\documentclass[prc,twocolumn,showpacs,nofootinbib]{revtex4}

\usepackage{graphicx}
\begin{document}
\bibliographystyle{apsrev}
 
\title{A Time Dependent Local Isospin  Density Approximation Study of 
Asymmetric Nuclear Matter}

\author{Enrico Lipparini}
\email{lipparin@science.unitn.it}
\affiliation{Dipartimento di Fisica, University of Trento, via Sommarive 14, 
I--38123,
Povo, Trento, Italy}
\affiliation{INFN-TIFPA, Trento Institute of Fundamental Physics and Applications, Trento, Italy}

\author{Francesco Pederiva}
\email{pederiva@science.unitn.it}
\affiliation{Dipartimento di Fisica, University of Trento, via Sommarive 14,
I--38123,
Povo, Trento, Italy}
\affiliation{INFN-TIFPA, Trento Institute of Fundamental Physics and Applications, Trento, Italy}

\date{\today}

\begin{abstract} 
The dynamic response of asymmetric nuclear matter is studied by using a Time-Dependent Local Isospin Density (TDLIDA) approximation
approach. Calculations are based on a local density energy functional derived by an Auxiliary Field Diffusion Monte Carlo (AFDMC) calculation of bulk nuclear matter.
Three types of excited states emerge: collective states, a continuum of quasi-particle-quasi-hole excitations and
unstable solutions. These states are analyzed and discussed for different values of the nuclear density $\rho$ and isospin asymmetry
$\xi=(N-Z)/A$. An analytical expression of the compressibility as a function of $\rho$ and $\xi$ is derived which
show explicitly an instability of the neutron matter around $\rho\simeq 0.09 fm^{-3}$ when a small fraction of protons is
added to the system.

\end{abstract}


\maketitle

\section{Introduction}
The dynamic response function and the dynamic structure factor of nuclear matter at $\beta$-equilibrium are key ingredients to understand the mechanisms related to interaction processes occurring in the interior of neutron stars. 
As it is well known since the seminal works of Sawyer\cite{Sawyer75, Sawyer89}, the structure factor is related to the mean free path of neutrinos in the nuclear medium\cite{Iwamoto82}. The opacity of nuclear matter to neutrinos, and its implication in the neutrino emission in the early phases of existence of a neutron star have been widely reviewed in the paper of Burrows et al. \cite{Burrows06}. 

Another interesting piece of information that can be extracted from the knowledge of the dynamic structure factor concerns the mechanical instability that can be related to divergencies in the response function, and in a nonphysical negative value of the compressibility. This is particularly interesting at values of the density where homogeneous matter is supposed to give way to more exotic inhomogeneous phases, before reaching the regime in which neutron rich nuclei are dominant\cite{Chamel08, Burrows06}. An interesting attempt to compute the structure factor of neutron matter by means of semiclassical simulations in this regime was recently performed by Horowitz et al.\cite{Horowitz04,Horowitz05}.

In general, several calculations of the low momentum structure factor of neutron matter have been performed using different methods, for instance in the framework of BHF-RPA theories, both at zero and finite temperature\cite{Shen03}, or in the Landau Theory approach starting from CBF effective interactions\cite{Benhar13}. 

In this paper we propose a different route to the computation of the dynamic structure factor in the isoscalar and isovector channels for an infinite nuclear matter with an arbitrary asymmetry. 

The approach is based on first deriving a reasonable isospin-density functional from an equation of state (EoS) computed by means of the Auxiliary Field Diffusion Monte Carlo (AFDMC)\cite{Schmidt99} method by Gandolfi et al.\cite{G10}. In that paper the EoS is derived from a density-dependent interaction effectively in the including many-body contributions to the interaction. This Hamiltonian was fitted to correctly reproduce the saturation properties of symmetric nuclear matter. However, the predictions of the EoS of pure neutron matter and matter at $\beta$-equilibrium, including both electrons and muons, produce realistic values of quantities such as the mass/radius relation in the same range of accuracy of i.e. the APR EoS\cite{Akmal98}.

Obviously the range of densities that can be safely investigated relying on this EoS are limited. In the low density limit nuclear matter becomes unstable, and neutron matter should be described including pairing correlations\cite{Gandolfi09}. At densities of order 2$\rho_0$, where $\rho_0$=0.16fm$^{-3}$ is the nuclear matter saturation density, one can expect the onset of hypernuclear degrees of freedom\cite{Dapo10}, which completely changes the scenario. 

The isospin-density functional is then used in a generalized linear response framework based on the Time Dependent Local Density approximation (TDLDA). Following the idea of the Local Spin Density Approximation (LSDA, see e.g.\cite{Perdew86}), and its extension to the study of the dynamics of the system\cite{Serra97}, we derive the response function in a Time Dependent Local Isospin Approximation (TDLIDA), and compute all the relevant physical quantities such as the frequency and strength of the collective modes and the compressibility for various values of the density and of the asymmetry at T=0.

The paper is organized as follows. In Section II we will introduce in detail the derivation of the energy density functional; in Section II the extension of the TDLDA to the case of excitation in the isovector channel is discussed; Section IV contains the description of the resuts, and Section IV is devoted to conclusions.

\section{Derivation of the energy density functional}

By following the Kohn-Sham method\cite{Ko65}, we introduce the local isospin density approximation (LIDA) for nuclear matter defining the energy functional as:
\begin{eqnarray}
\label{eq 1a}
E(\rho,\xi)= T_0(\rho,\xi) + \int~\epsilon_V(\rho,\xi)~\rho ~d{\bf r} 
\end{eqnarray}
where $T_0(\rho,\xi)$ is the kinetic energy of the non interacting system with density  $\rho({\bf r})=\rho_n({\bf r})+\rho_p({\bf r})$, and 
isospin polarization $\xi=\rho_1/\rho$ with $\rho_1({\bf r})=\rho_n({\bf r})-\rho_p({\bf r})$, and where $\rho_n$ and $\rho_p$ are the neutron and proton densities, respectively.
Note that $T_0$ is the kinetic energy of the non-interacting system, therefore missing the contribtions coming from dynamical quantum correlations. However, it has a comparable magnitude and is consistently treated in this method.
Equation (\ref{eq 1a}) defines the interaction-correlation energy per particle $\epsilon_V(\rho,\xi)$ of the asymmetric nuclear matter with density
$\rho$ and isospin polarization $\xi$. 
This quantity can be extracted by any independent calculation
of the total energy per particle of asymmetric nuclear matter after subtraction of the free kinetic energy contributions at each value of the  $\rho$  and $\xi$.

In this work  we will use for $\epsilon_V(\rho,\xi)$  
the following parametrization based on 
the Auxiliary Field Diffusion Monte Carlo (AFDMC) calculations of Ref. \onlinecite{G10}: 
\begin{eqnarray}
\label{eq 1d} 
\epsilon_V(\rho,\xi)=\epsilon_0(\rho) + \xi^2\left[\epsilon_1(\rho) - \epsilon_0(\rho)\right]~,
\end{eqnarray}   
where
\begin{equation}
\label{eq 33}
\epsilon_q(\rho)= \epsilon^0_q +a_q(\rho-\rho_0)
+b_q(\rho-\rho_0)^2+c_q(\rho-\rho_0)^3 e^{\gamma_q(\rho-\rho_0)}~,
\end{equation}
are the interaction correlation energies of symmetric nuclear ($\xi=0,q=0$), and neutron matter ($\xi=1,q=1$), respectively. We have assumed the saturation density value $\rho_0=0.16$fm$^{-3}$.
The values of the parameters in Eq. (\ref{eq 33}) we have extracted from the fit 
are given in Table I.
\begin{table}
\begin{ruledtabular}
\begin{tabular}{cccccc}
$q$&$\epsilon_q$ &$a_q$ &$b_q$ &$c_q$
 &$\gamma_q$ \\
&(MeV)&(MeV$\cdot$fm$^3$)&(MeV$\cdot$fm$^6$)&(MeV$\cdot$fm$^9$)&(fm$^3$)\\
\hline
0& -38.1 & -92.1 & 630.1 & -1717.2 & -2.360\\
1& -19.8 & -21.0 & 533.0 & -1327.7 & -2.201\\
\end{tabular}
\end{ruledtabular}
\label{table_coef}
\caption{Coefficients of the parameterization of the interaction-correlation energy of symmetric nuclear matter ($q$=0) and pure neutron matter ($q$=1) derived by fitting the equation of state of Ref. \cite{G10}.}
\end{table}

The parametrization (\ref{eq 1d}) reproduces very well  the AFDMC calculations  in a wide range
of density $\rho$ (from $\rho_0/2$ to $3\rho_0$) and polarizations $\xi$.

By minimizing the energy functional (\ref{eq 1a}) with the constraint that the number of neutrons and protons
remains constant,  we can derive a set of self-consistent equations by exactly treating the  kinetic energy functional $T_0$. 
The detailed derivation can be found in Ref.\cite{Lip08}. The resulting set of coupled self-consistent equations for neutron and protons wave functions, assuming $\hbar=c=1$, 
are given by:
\begin{eqnarray}
\label{eq 1}
\left[\rule{0cm}{0.5cm}\right.
-\frac{1}{2m} \nabla_{\bf r}^2 
+v({\bf r}) + w({\bf r})\eta_{\tau}
\left.\rule{0cm}{0.5cm}\right]\varphi^{\tau}_i({\bf r})
=\varepsilon_{i,\tau}\,\varphi^{\tau}_i({\bf r})\; ,
\end{eqnarray}
where $i$ stands for the set of quantum numbers, excluding isospin,
that characterize the single-particle wave functions. 
The nuclear neutron and proton densities are given by 
$$
\rho_{\tau}=\sum_{i}\vert\varphi^{\tau}_i({\bf r})\vert^2
$$
with
$\eta_{\tau}= 1(-1)$ if $\tau=n (p)$ 
and $n, p$ stands 
for neutrons and protons, respectively. The effective potentials in Eq $\ref{eq 1}$ 
are derived from the interaction-correlation energy functional deriving with respect to the density and isospin polarization:
\begin{equation}
\label{eq 2}
v({\bf r})=\frac{\partial   \rho\epsilon_V[\rho({\bf r}),\xi]}{\partial\rho({\bf r})}~~~~ 
,
~~~~ w({\bf r})=\frac{\partial\epsilon_V[\rho({\bf r}),\xi)]}{\partial \xi({\bf r})}~.
\end{equation}

As it is well known, the Kohn-Sham method sketched here 
gives an exact solution of the variational principle which minimizes 
the energy functional (\ref{eq 1a}). The theory is still approximate in the sense
that the exact interaction-energy functional is unknown, and one needs to
rely on some expression derived by independent calculations. 
Concerning our choice for $\epsilon_V(\rho,\xi)$, we want to point out once more 
that, 
as discussed in Ref.\onlinecite{G10}, the functional (\ref{eq 1a})
gives an equation of state of asymmetric nuclear matter providing realistic predictions for neutron stars properties when the $\beta$-equilibrium condition is imposed. We can therefore be confident that both the density and isospin polarization dependence are sufficiently accurate to yeld reasonable values also
for the first and second derivatives of $\epsilon_V$ with 
respect to $\rho$ and $\xi$. These are the main ingredients of the
Kohn-Sham method and of its time-dependent version described in the next section. 

Finally, we notice that when applied to infinite nuclear matter, the static equations (\ref{eq 1}) are satisfied by plane-wave solutions for $\varphi^{\tau}_i({\bf r})$, since in this case all the densities are constant as well as the density dependent potentials (\ref{eq 2}). 
As a consequence, the Kohn-Sham equilibrium density and energy per particle are obviously the same of the starting interaction used to derive the energy functional. 
Hence, the static Kohn-Sham equations do not give any information in infinite systems.
However, as described in the next section, the {\it time-dependent} version of them give new and useful solutions for the nuclear response even starting from the same simple ingredients.

\section{TDLIDA response of infinite asymmetric nuclear matter}

In this section we calculate the TDLIDA density-density ($\chi^s(q,\omega)$) and isovector-density/isovector-density ($\chi^v(q,\omega)$) 
response functions of a 3-dimensional spin-unpolarized  
uniform gas of $N$ neutrons and $Z$ protons ($N+Z=A$), with isospin polarization $\xi=\rho_1/\rho$.

We start writing the time-dependent Kohn-Sham (KS) equations in an external, time-dependent field along the {\bf r}-direction:
\begin{equation}
\label{eq 3a}      
\sum _{k=1}^{A}\lambda^{k}_{\tau}(e^{i ({\bf q}\cdot{\bf r}_{k}-\omega 
t)}+e^{-i({\bf q} 
\cdot{\bf r}_{k}-\omega t)})~~,
\end{equation}
with  $\lambda^{k}_{\tau}=\lambda$ for density excitations and 
$\lambda^{k}_{\tau}=\lambda\eta_{\tau}$ for isovector-density excitations. The KS equations read:
\begin{eqnarray}
\begin{array}{rcl}
i\frac{\partial }{\partial t}\varphi^{\tau}_i\left({\bf r},t\right)
&=&\left\{-\frac{1}{2m} \nabla_{\bf r}^{2} +
+v[\rho_n\left({\bf 
r},t\right),\rho_p\left({\bf r},t\right)] 
\right.

\\ \\
&&\left. + 
w[\rho_n\left({\bf 
r},t\right),\rho_p\left({\bf r},t\right)]\eta_{\tau}
+\lambda_{\tau}\, [e^{i\left({\bf q}\cdot{\bf r} -\omega
t\right) }\right. \\ \\
&&
\left.+e^{-i\left( {\bf q}\cdot{\bf r}-\omega t\right) }] \right\}
\varphi^{\tau}_i\left({\bf r},t\right) \, .
\end{array}
\label{eq 3b}
\end{eqnarray}

In the uniform 3D nucleon gas, the nuclear density oscillations 
induced by the external field are given by:
\begin{eqnarray}
\label{eq 4a}
\rho_n \left({\bf r},t\right)&=& \rho_n + \delta 
\rho_n 
\left({\bf r},t\right) \; ,
\nonumber
\\
\rho_p \left({\bf r},t\right)&=& \rho_p + \delta 
\rho_p\left({\bf r},t\right) \; ,
\end{eqnarray}
where $\rho_n$ and $\rho_p$ are the neutron  and proton constant densities 
of the unperturbed initial state, respectively, and
\begin{eqnarray}
\delta \rho_n\left({\bf r},t\right)&=&\delta\rho_n\left( 
e^{i\left({\bf q}\cdot{\bf r}-\omega t\right) }
+e^{-i\left({\bf q}\cdot{\bf r}-\omega t\right) }\right)~,
\nonumber
\\
\delta \rho_p \left({\bf r},t\right)&=&\delta \rho_p\left( 
e^{i\left({\bf q}\cdot{\bf r}-\omega t\right) }
+e^{-i\left({\bf q}\cdot{\bf r}-\omega t\right) }\right)~,
\label{eq 4b}
\end{eqnarray}
as follows from translational invariance. The quantities $\delta\rho_n$ and
$\delta\rho_p $  are   constants to be determined.
Eqs. (\ref{eq 3b}) and (\ref{eq 4b}) have solutions describing density fluctuations 
in the density operator $F=\sum ^{A}_{k=1}e^{i {\bf q}\cdot{\bf r}_{k}}$,  and 
isovector-density operator
$F_{\tau}=\sum ^{A}_{k=1}e^{i {\bf q}\cdot{\bf r}_{k}}\eta^k_{\tau}$, given by
\begin{eqnarray}
\delta F(\hat{O},\omega)=\langle\psi(t)|F|\psi(t)\rangle
-\langle 0|F|0\rangle=\nonumber\\=
\int d {\bf r} \,e^{i {\bf q}\cdot{\bf r}}[ \rho({\bf r},t)
-\rho]_{\hat{O}}= V \,e^{i\omega 
t}(\delta\rho_n+\delta\rho_p) ~,
\label{eq 5a}
\end{eqnarray}
and,
\begin{eqnarray}
\begin{array}{l}
\delta F_{\tau}(\hat{O},\omega)=\langle\psi(t)|F_{\tau}|\psi(t)\rangle
-\langle 0|F_{\tau}|0\rangle=\\ \\=
\displaystyle
\int d {\bf r} \,e^{i {\bf q}\cdot{\bf r}}[ \rho_1({\bf r},t)
-\rho_{1}]_{\hat{O}}=\\ \\
= V \,e^{i\omega 
t}(\delta\rho_n-\delta\rho_p)~,
\end{array}
\label{eq 5b}
\end{eqnarray}
where  
$V$ is the volume of the gas and $\hat{O}=\sum ^{A}_{k=1}e^{-i {\bf 
q}\cdot{\bf r}_{k}}$. 

The density-density response is given by\cite{Lip08}
\begin{equation}
\chi^{s}(q,\omega)=\frac{V(\delta\rho_n+\delta\rho_p)}{\lambda }
\equiv\chi^n(q,\omega)+\chi^p(q,\omega) ~,
\label{eq 6a}
\end{equation}
and the isovector-density-isovector-density response is:
\begin{equation}
\chi^{v}(q,\omega)=\frac{V(\delta\rho_n-\delta\rho_p) }{\lambda }
\equiv\chi^n(q,\omega)-\chi^p(q,\omega) ~.
\label{eq 6b}
\end{equation}

In order to determine $\delta\rho_n$ and $\delta\rho_p$, 
we then insert $\rho_n({\bf r},t)$,$\rho_p({\bf r},t)$ of Eqs. (\ref{eq 4a})-(\ref{eq 4b}) into
(\ref{eq 3b})
and linearize the equations. This procedure determines the self-consistent
KS mean-field potential entering Eq. (\ref{eq 3b}) to be:
\begin{eqnarray}
\begin{array}{rcl}
V_{KS}[\rho_n({\bf r} 
,t),\rho_p({\bf r},t)]&=&V_{KS}(\rho_n,\rho_p)+
\\ &+&\left. \frac{\partial V_{KS}}{\partial\rho_n({\bf r},t)}
\right|_{\rho_n,\rho_p}
\delta \rho_n({\bf r},t) 
\\ &+& \left. \frac{\partial V_{KS}}{\partial\rho_p({\bf r},t)}
\right|_{\rho_n,\rho_p}\delta 
\rho_p({\bf r},t)~.
\label{eq 7}
\end{array}
\end{eqnarray}
Evaentually, from Eqs. (\ref{eq 3b}) and (\ref{eq 7}) we obtain:
\begin{widetext}
\begin{eqnarray}
\begin{array}{c}
\displaystyle
i\frac{\partial }{\partial t}\varphi^n_i\left({\bf r},t\right)=
\left\{-\frac{1}{2m}\nabla_{\bf r}^{2} + const.
  +[\delta \rho_n V_{n,n}+ \delta \rho_p V_{n,p}+ \lambda]\, 
\left( e^{i\left( {\bf q}\cdot{\bf r}-\omega
t\right) }
+e^{-i\left( {\bf q}\cdot{\bf r}-\omega t\right) }\right) \right\}
\varphi^n_i\left({\bf r},t\right) \, ,
\\ \\
\displaystyle
i\frac{\partial }{\partial t}\varphi^p_i\left({\bf r},t\right)=
\left\{-\frac{1}{2m}\nabla_{\bf r}^{2} + const.
+[\delta \rho_n V_{n,p}+ \delta \rho_p V_{n,n}
\pm\lambda]\, \left( e^{i\left( {\bf q}\cdot{\bf r}-\omega
t\right) }
+e^{-i\left( {\bf q}\cdot{\bf r}-\omega t\right) }\right) \right\}
\varphi^p_i\left({\bf r},t\right) \, 
\end{array}
\end{eqnarray}
\label{eq 8}
\end{widetext}
where  in the second equation
one must keep the plus sign when calculating the density-density response and the
minus sign when calculating the isovector-density/isovector-density response.
In Eqs. (\ref{eq 8}) we have defined the mean-field potentials:
\begin{eqnarray}
\begin{array}{l}
V_{nn}=\left. \frac{\partial (v+w)}{\partial\rho_n({\bf r},t)}
\right|_{\rho_n,\rho_p}~=~\left.\left(\frac{\partial}{\partial\rho}+{1\over\rho}\frac{\partial}{\partial\xi}\right)\left(v+w\right)\right|_{\rho,\xi},
\\ \\
V_{np}=\left. \frac{\partial (v+w)}{\partial\rho_p({\bf r},t)}
\right|_{\rho_n,\rho_p}~=~\left.\left(\frac{\partial}{\partial\rho}-{1\over\rho}\frac{\partial}{\partial\xi}\right)\left(v+w\right)\right|_{\rho,\xi},
\\ \\
V_{pn}=\left. \frac{\partial (v-w)}{\partial\rho_n({\bf r},t)}
\right|_{\rho_n,\rho_p}~=~\left.\left(\frac{\partial}{\partial\rho}+{1\over\rho}\frac{\partial}{\partial\xi}\right)\left(v-w\right)\right|_{\rho,\xi},
\\ \\
V_{pp}=\left. \frac{\partial (v-w)}{\partial\rho_p({\bf r},t)}
\right|_{\rho_n,\rho_p}~=~\left.\left(\frac{\partial}{\partial\rho}-{1\over\rho}\frac{\partial}{\partial\xi}\right)\left(v-w\right)\right|_{\rho,\xi}.
\end{array}
\label{eq 8b}
\end{eqnarray}

Eq. (\ref{eq 8}) can be rewritten as
\begin{eqnarray}
\begin{array}{l}
\displaystyle
i\frac{\partial }{\partial t}\varphi^n_i\left({\bf r},t\right)=\\ \\= \displaystyle
\left\{-\frac{1}{2m}\nabla_{\bf r}^{2} + C
+\lambda_{n}^{\prime}\, [ e^{i\left( {\bf q}\cdot{\bf r}-\omega
t\right) }
+e^{-i\left({\bf q}\cdot{\bf r} -\omega t\right) }] \right\}
\varphi^n_i\left({\bf r},t\right)
\\ \\
\displaystyle
i\frac{\partial }{\partial t}\varphi^p_i\left({\bf r},t\right)=\\ \\= \displaystyle
\left\{-\frac{1}{2m}\nabla_{\bf r}^{2} + C
+\lambda_{p}^{\prime}\, [ e^{i\left( {\bf q}\cdot{\bf r}-\omega
t\right) }
+e^{-i\left( {\bf q}\cdot{\bf r}-\omega t\right) }] \right\}
\varphi^p_i\left({\bf r},t\right)
\end{array}
\label{eq 8c}
\end{eqnarray}
where $C$ is a constant, and
\begin{eqnarray}
\lambda_{n}^{\prime}= \delta \rho_n V_{n,n}+ \delta 
\rho_p V_{n,p}+\lambda~~,
\nonumber
\\
\lambda_{p}^{\prime}=\delta \rho_n V_{p,n}+ \delta 
\rho_p V_{p,p}
\pm\lambda ~~.
\label{eq 9}
\end{eqnarray}
Equation
(\ref{eq 8c}) coincides with that of a non-interacting system coupled to an
external time oscillating field, with a coupling constant $\lambda
^{\prime }$
given by Eq. (\ref{eq 9}). For such a system, the density response
functions are
the single-particle free responses $\chi^n_0(q,\omega)$, $\chi^p_0(q,\omega)$.   
From Eqs. (\ref{eq 6a}) and (\ref{eq 6b})
and from
the analogous relations for the free response functions
\begin{eqnarray}
\chi^n_0(q,\omega)=\frac{V \delta \rho_n }{\lambda_n ^{\prime } } ~,
\nonumber
\\
\chi^p_0(q,\omega)=\frac{V \delta \rho_p }{\lambda_p ^{\prime } }
\label{eq 9b}
\end{eqnarray}
we  obtain
\begin{eqnarray}
\lambda\,\chi^n(q,\omega)=
\lambda_{n} ^{\prime }\,\chi^n_0(q,\omega)=L \delta\rho_n~,\nonumber
\\ \\
\lambda\,\chi^p(q,\omega)=
\lambda_p ^{\prime }\,\chi^p_0(q,\omega)=L \delta\rho_p~.\nonumber
\label{eq 9c}
\end{eqnarray}
The solution of these equations, finally gives
the TDLIDA response functions
\begin{widetext}
\begin{eqnarray}
\begin{array}{c}
\displaystyle
\chi^{s}=V\frac{\chi^n_0[V-(V_{p,p}-V_{n,p})
\chi^p_0]
+\chi^p_0[V-(V_{n,n}-V_{p,n})\chi^n_0]}
{(V-V_{p,p}
\chi^p_0)(V-V_{n.n}\chi^n_0])
-V_{n,p}\chi^n_0V_{p,n} \chi^p_0} ~,
\\ \\
\displaystyle
\chi^{v}=V\frac{\chi^n_0[V-(V_{p,p}+V_{n,p})
\chi^p_0]
+\chi^p_0[V-(V_{n,n}+V_{p,n})\chi^n_0] }
{(V-V_{p,p}
\chi^p_0)(V-V_{n,n}\chi^n_0)
-V_{n,p}\chi^n_0V_{p,n} \chi^p_0}~~.
\end{array}
\label{eq 10}
\end{eqnarray}
\end{widetext}

Eqs. (\ref{eq 10}) allow for studying the response of partially isospin-polarized nuclear matter ($N\ne Z$), which is the aim of the present work. 

Note that for fully isospin-unpolarized systems (N=Z)  the 
above equations are drastically simplified. In fact, in this case, one has:
$\rho_n=\rho_p=\rho/2$,  
$\chi^n_0=\chi^p_0=\chi_0/2$ and 
$V_{n,n}=V_{p,p}=
\left. (\frac{\partial v}{\partial\rho}+{1\over\rho}\frac{\partial w}{\partial \xi})
\right|_{\rho,\xi=0}$, $V_{n,p}=V_{p,n}=
\left. (\frac{\partial v}{\partial\rho}-{1\over\rho}\frac{\partial w}{\partial \xi})
\right|_{\rho,\xi=0}$
and one gets from Eqs.  (\ref{eq 10})
\begin{equation}
\chi^{s}(q,\omega)={\chi_0(q,\omega)\over 1 -\left. \frac{\partial v}{\partial\rho}
\right|_{\rho,\xi=0}
{\chi_0(q,\omega)\over V}} ~~~,
\label{eq 11a}
\end{equation}
and
\begin{equation}
\chi^{v}(q,\omega)={\chi_0(q,\omega)\over 1 -\left. {1\over\rho}\frac{\partial
w}{\partial \xi}
\right|_{\rho,\xi=0}
{\chi_0(q,\omega)\over V}} ~~~,
\label{eq 11b}
\end{equation}
which have the same form of the RPA responses of isospin unpolarized systems, but not the same meaning, since they have as main ingredients the
derivative of the self-consistent KS potentials, and not the Fourier transform of an effective nucleon-nucleon interaction as in the RPA theory. 

The same simplifications occurs in the case of pure neutron matter. In this case one finds that $\chi^p_0=0$. Consequently:
\begin{equation}
\chi^{s}(q,\omega)=\chi^{v}(q,\omega)={\chi^n_0(q,\omega)\over 1 - V_{n,n}{\chi^n_0(q,\omega)\over V}} ~~~.
\label{eq 11c}
\end{equation}

Since the time-dependent-density functional approach only holds in the low-$q$, low-$\omega$ limits\cite{Lip08},
in the following we will use as the free response functions $\chi^n_0$ and $\chi^p_0$ entering Eqs. (\ref{eq 10}) 
the following simple expressions valid in these limits:
\begin{eqnarray}
\begin{array}{l}
\chi^{n,p}_0({\bf q},\omega)=\\ \\=-V~\nu^{n,p}
\left[1+\frac{s}{2(1\pm\xi)^{1/3}}\ln {s -(1\pm\xi)^{1/3})\over s+(1\pm\xi)^{1/3})}  \right],
\end{array}
\label{eq 11d}
\end{eqnarray}
where $\nu^{n,p}=m k_F^{n,p}/\pi^2=m k_F(1\pm\xi)^{1/3}/\pi^2$,
$k_F=(\frac{3\pi^2}{2}\rho)^{1/3}$, $s=\omega/(q v_f)$. The plus sign holds for $\chi^{n}_0$ and the minus sign for $\chi^{p}_0$.   
Note that in the low-$q$, low-$\omega$ limits, $\chi^{n}_0$ and  $\chi^{p}_0$ depend on $q$ and $\omega$
only through the combination $s=\omega/(q v_f)$. Since $\chi^{s,v}$ of Eqs. (\ref{eq 10}) depends on $q$ and $\omega$ only via the dependence of 
 $\chi^{n}_0$ and  
$\chi^{p}_0$ on these variables,  
also the interacting responses turn out to be functions of $s$ only.
As in Landau theory, also in TDLIDA the nuclear responses 
are functions of $s=\omega/(q v_f)$ and not of $q$ and $\omega$ separately.  

\section{Results}

\subsection{Mean-field potential}
From equations (\ref{eq 10}), by taking the immaginary part of $\chi$, it is possible to calculate the excitation strength 
$S^{s,v}(q,\omega)=-(1/\pi) Im~\chi^{s,v}$ and the moments $m^{s,v}_k$ of  the density ($s$) and isovector density ($v$) excitation operators $F^{s,v}$, with $F^s=\sum ^{A}_{k=1}e^{i {\bf 
q}\cdot{\bf r}_{k}}$ 
and $F^{v}=\sum ^{A}_{k=1}e^{i {\bf q}\cdot{\bf r}_{k}}\eta^k_{\tau}$. The moments are given by:
\begin{eqnarray}        
m^{s,v}_{k}=\int_0^{\infty} d\omega\; \omega ^{k}S^{s,v}(q,\omega)=\sum _{n}\omega ^{k}_{no}| \langle 0| F^{s,v}| n\rangle | ^{2}\,.
\label{eq 15}
\end{eqnarray}
By setting the mean-field potentials $V_{nn}$, $V_{pp}$, $V_{np}$  equal to zero it is possible to write:
\begin{eqnarray}
\begin{array}{l}
S_{free}^{s,v}(q,\omega)=\\ \\={V m k_F\over\pi^2}{s\over2}\left[\Theta\left(1-{s\over(1+\xi)^{1/3}}\right)
+\Theta\left(1-{s\over(1-\xi)^{1/3}}\right)\right]
\end{array}
\label{eq 111}
\end{eqnarray}
which gives the one-particle-one-hole excitation strength, and, by integration, the Fermi gas energy moments of the asymmetric non interacting nuclear matter. 
\begin{figure}

\centerline{\includegraphics[scale=0.36]{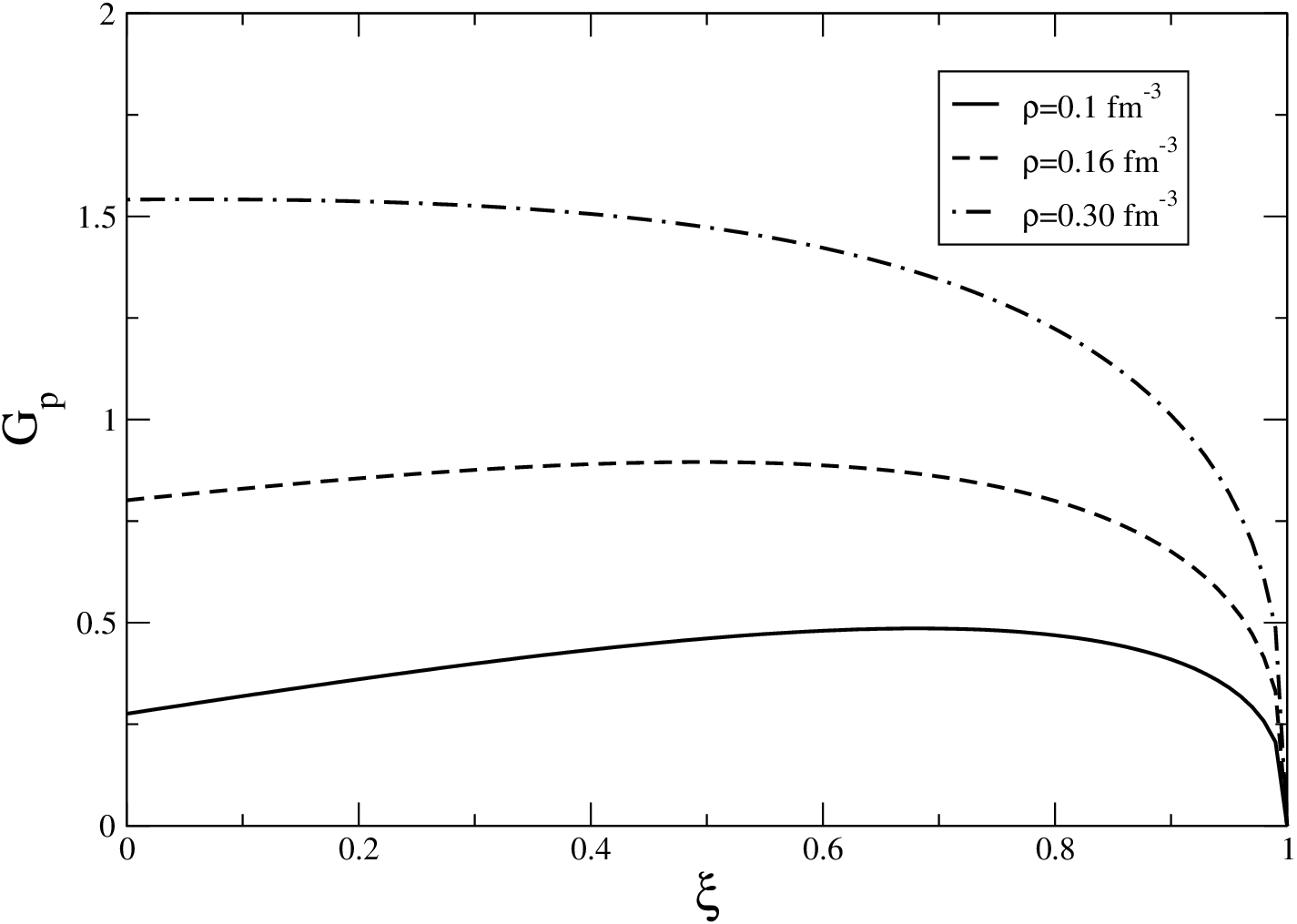}}

\caption[]{Adymensional interaction parameter $G_p=\nu_p V_{pp}$ as a 
function of the isospin asymmetry $\xi=(N-Z)/A$ for three different values of the density $\rho$.}
\label{fig1}
\end{figure}

\begin{figure}

\centerline{\includegraphics[scale=0.36]{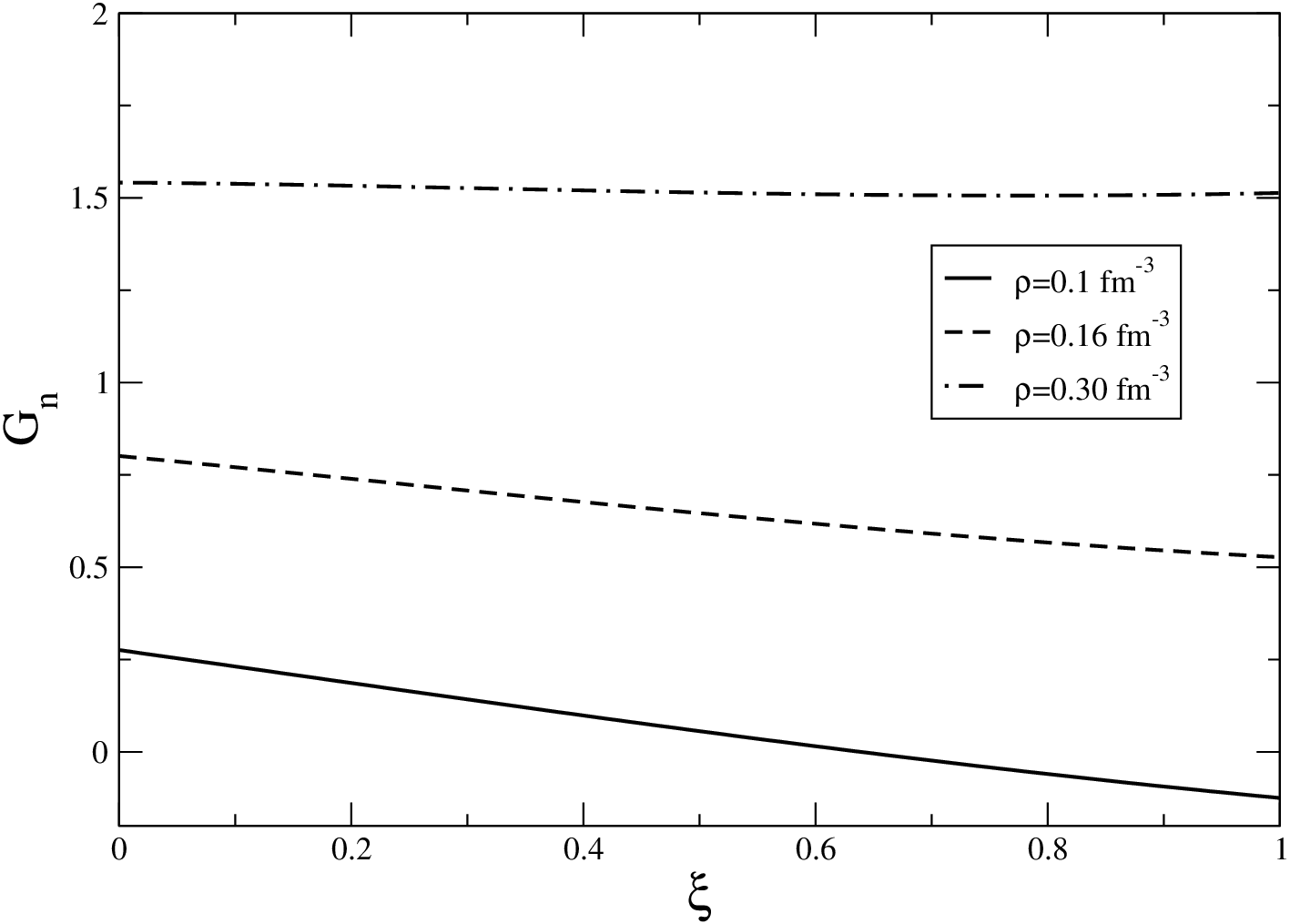}}

\caption[]{Same as in Fig. 1 for the interaction parameter $G_n=\nu_n V_{nn}$.}
\label{fig2}
\end{figure}

\begin{figure}

\vspace{1cm}

\centerline{\includegraphics[scale=0.36]{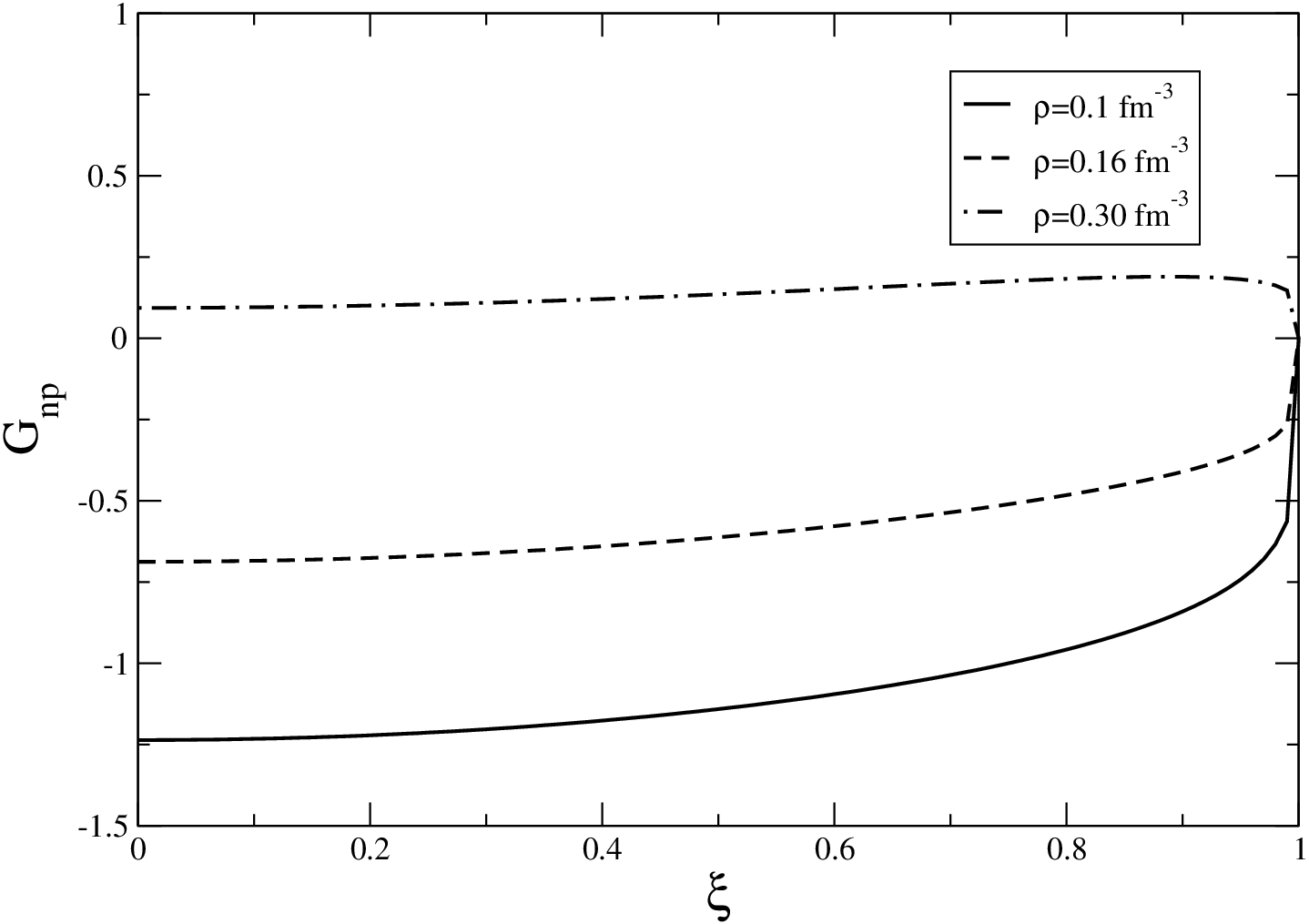}}

\caption[]{Same as in Fig. 1 for the interaction parameter $G_{np}=\sqrt{\nu_{n}\nu_{p}} V_{np}$.}
\label{fig3}
\end{figure}

In Figs. 1-3 and Tables II-IV the values  
of the adimensional interaction parameters $G_n=\nu_n V_{nn}$, $G_p=\nu_p V_{pp}$ and 
$G_{np}=\sqrt{\nu_n\nu_p} V_{np}$ are reported as a function of the isovector 
polarization 
$\xi=(N-Z)/A$ for three values of the density $\rho$. 
These quantities show a strong $\xi$ and $\rho$ dependence,
giving rise to excitations strengths and moments which are quite
different from the non interacting cases.

\subsection{Energy weighted sum rules}
We first consider the moments $m_{-1}, m_1, m_3$ of the density ($s$) and isovector density ($v$) operators, $F^s$ and $F^v$.
These quantities  still have an analytical expressions, given by:
\begin{equation}   
m^{s,v}_{-1}={V\over2}{\nu_n(1+G_p)+\nu_p(1+G_n) \mp 2\sqrt{\nu_p\nu_n}G_{np}
\over (1+G_p)(1+G_n)-G_{np}^2}\,,\label{eq 16}
\end{equation}
\begin{equation}
m^{s,v}_{1}={N\over2}{q^2\over m}\,,\label{eq 17}
\end{equation}
\begin{eqnarray}
\begin{array}{lcl}
\displaystyle
m^{s,v}_{3}&=&{V\over2}{q^4 \over {m}^8}\left\{{1\over5}\left(\nu_n^5+\nu_p^5\right)\right.\\ \\ 
\displaystyle
&+&\left. {1\over9}\left[\nu_n^5G_n+\nu_p^5G_p\pm2\nu_n^{5/2}\nu_p^{5/2}G_{np}
\right]\right\},
\end{array}
\label{eq 18}
\end{eqnarray}
where  the minus sign applies to density excitations  and the plus sign to the isovector-density ones.

The $m_{-1}$ moment is known as the {\it hydrodynamic sum
rule} because in the $N=Z$ case it is directly related to
ordinary sound velocity. The $m_1$ moment is the $f$-sum  
rule, and the $m_3$ moment is related to the elastic properties of the
Fermi systems (for a wide illustration of the properties of these sum rules see Ref.\onlinecite{Lip08}).
In the case of pure neutron matter, the  $m_{-1}$ moment is related to the neutron matter compressibilty $K^n$ by the relation:
\begin{eqnarray}
{m_{-1}\over m^0_{-1}}={K^n\over K^{n}_0}={1\over 1+G_n},
\label{eq 19b}
\end{eqnarray}
where $m^0_{-1}=V \nu_n/2$ and $K^n_0=9\pi^2m/k{^n}_F^5$ respectively are the Fermi gas static polarizability and compressibility.
Similar expressions also hold in the symmetric nuclear matter ($N=Z$), where $\nu_n=\nu_p$, and $G_p=G_n=G$. In this case the moments read 
\begin{eqnarray}
{m^s_{-1}\over m^0_{-1}}={K\over K_0}={1\over 1+G+G_{np}},
\label{eq 19c}
\end{eqnarray}
and
\begin{eqnarray}
{m^v_{-1}\over m^0_{-1}}={1\over 1+G-G_{np}}~~,
\label{eq 19d}
\end{eqnarray}
for the isoscalar and isovector cases, respectively. Expressions (\ref{eq 19b},\ref{eq 19c}) 
are quite similar to the results of Landau theory for neutron and 
symmetric nuclear matter\cite{Lip08}.
The true novelty of Eqs. (\ref{eq 10}) and (\ref{eq 16},\ref{eq 17},\ref{eq 18}) stands in their {\it explicit}  dependence on $N$, $Z$ which allows to study the properties of the nuclear response when a fraction of protons is present in the neutron matter as happens in the neutron star interior.   

Note that in the local isospin density approximation used here, the $f$-sum rule $m_1$ is the same for both the isoscalar and
isovector excitation operators, whereas calculating $m_1^v$ directly from the original interaction used to deriving the local energy functional,
woud have yelded for $m_1^v$ an interaction  contribution. 
This reflects a failure of the model in reproducing quantities in
the high $\omega$ region.
The sum rules (\ref{eq 16},\ref{eq 17},\ref{eq 18}) are only valid in the low-$q$ and  low-$\omega$ limits,
where the TDLIDA  is expected to work. They only account for one-particle-one-hole and collective excitations. Many-particle-many-hole excitations are 
important in the high energy part of the excitation spectrum \cite{Dal89}, and give an important contribution to $m_1^v$. 
Following this argument, since the $m_{-1}$ moment is mainly determined by the low energy part of the spectrum, one expects that this sum rule should 
be the closest to the the exact value that might in principle be computed by directly solving the many-body Schroedinger equation.

\subsection{Collective modes and dynamic structure factors}
We now turn to the interacting excitation strength $S(s=\omega/q v_F)$. The interaction produces new types of excitations beyond the usual
one-particle-one-hole ones which are the only excitations of the non interacting case. These excitations are given by the poles of the response 
functions (\ref{eq 10}), which are 
the solutions of the equations:
\begin{eqnarray}
(1+G_p
\Omega^p)(1+G_n\Omega^n)
-G^2_{n,p}\Omega^n\Omega^p=0~~,
\label{eq 20}
\end{eqnarray}
where $\Omega^{n,p}=\left[1+\frac{s}{2(1\pm\xi)^{1/3}}\ln {s -(1\pm\xi)^{1/3})\over s+(1\pm\xi)^{1/3})}  \right]$, with the 
plus sign holding for $\Omega^n$  and the minus sign for $\Omega^p$. 
Depending on the strength of the interaction, these new solutions are essentially of three types: a) real solutions such that
$s>(1\pm\xi)^{1/3}$ (collective modes), producing a discrete peak 
in the dynamic form factor $S(s)$ with no attenuation; b) Solutions with some imaginary component, and corresponding modes which 
 decay by exciting single quasi-particle-quasi-hole pairs (Landau damping); c) Unstable solutions which are associated to the divergence
of the polarizability sum rules $m^{s,v}_{-1}$. 

In Tables II-IV, we report the values ${\bar s}$ of $s$ at which the collective states occur in the isoscalar (${\bar s}_s$) and isovector 
 (${\bar s}_v$) strengths, the percentage $m_1^s$ and  $m_1^v$ of energy-weighted sum rule
exhausted by the collective states in the isoscalar and isovector channels, respectively, and  the two mean $s$ values 
\begin{eqnarray}
{\bar s}^{s,v}_{1,-1}={\omega^{s,v}_{1,-1}\over q v_F}=\sqrt{{m^{s,v}_{1}\over m^{s,v}_{-1}}}{1\over q v_F}\nonumber
\\
{\bar s}^{s,v}_{3,1}={\omega^{s,v}_{3,1}\over q v_F}=\sqrt{{m^{s,v}_{3}\over m^{s,v}_{1}}}{1\over q v_F},\label{eq 22}
\end{eqnarray}
obtained by Eqs. (\ref{eq 16},\ref{eq 17},\ref{eq 18}). The energy $\omega^{s,v}_{1,-1}$ is hereafter referred to as the {\it hydrodynamic} energy
in analogy to what happens in liquid He$^3$, where  $v_1=\omega^{s}_{1,-1}/q$ reproduces the predictions of the hydrodynamical model for the ordinary 
(first-) sound wave. 
Conversely the energy 
$\omega^{s,v}_{3,1}$ is referred to as  the {\it elastic} energy since in liquid He$^3$ $v_0=\omega^{s}_{3,1}/q$  reproduces the predictions 
of the elastic model for the zero-sound wave\cite{Lip08}.

The strength of the collective state at $s=\bar s$ can be computed from the following (adymensional) expression for the response:
\begin{widetext}
\begin{eqnarray} 
{\chi^{s,v}\over N m/(2 k_F^2)}=-3\frac{(1+\xi)^{1/3}\Omega^n[1+(G_p\mp ({(1-\xi\over1+\xi})^{1/6})G_{np})\Omega^p]
+(1-\xi)^{1/3}\Omega^p[1+(G_n\mp ({(1+\xi\over1-\xi})^{1/6})G_{np})\Omega^n]} 
{(1+G_p\Omega^p)(1+G_n\Omega^n)
-G^2_{n,p}\Omega^n\Omega^p} ~,
\label{eq 21}
\end{eqnarray}
\end{widetext}
derived from Eqs. (\ref{eq 10}) and (\ref{eq 11d}) by expanding
$\chi^{s,v}$ around the pole at $\bar s$. 
Naming $N(s)$ and $D(s)$ the numerator and the nenominator of Eq. (\ref{eq 21}), respectively, one obtains: 
\begin{equation}
\frac{S(s)}{Nm/(2 k_F^2)}=\frac{N(s)}{\displaystyle \frac{\partial D(s)}{\partial s}}\delta(s-\bar s). 
\end{equation}
The fraction of the $f$-sum rule $m_1$
exhausted by the collective state is easily calculated to be 
\begin{equation}
{\bar s}\frac{N(\bar s)}{\displaystyle \frac{\partial D(s)}{\partial s}|_{s=\bar s}}.
\end{equation}

\begin{figure}{t}

\centerline{\includegraphics[scale=0.34]{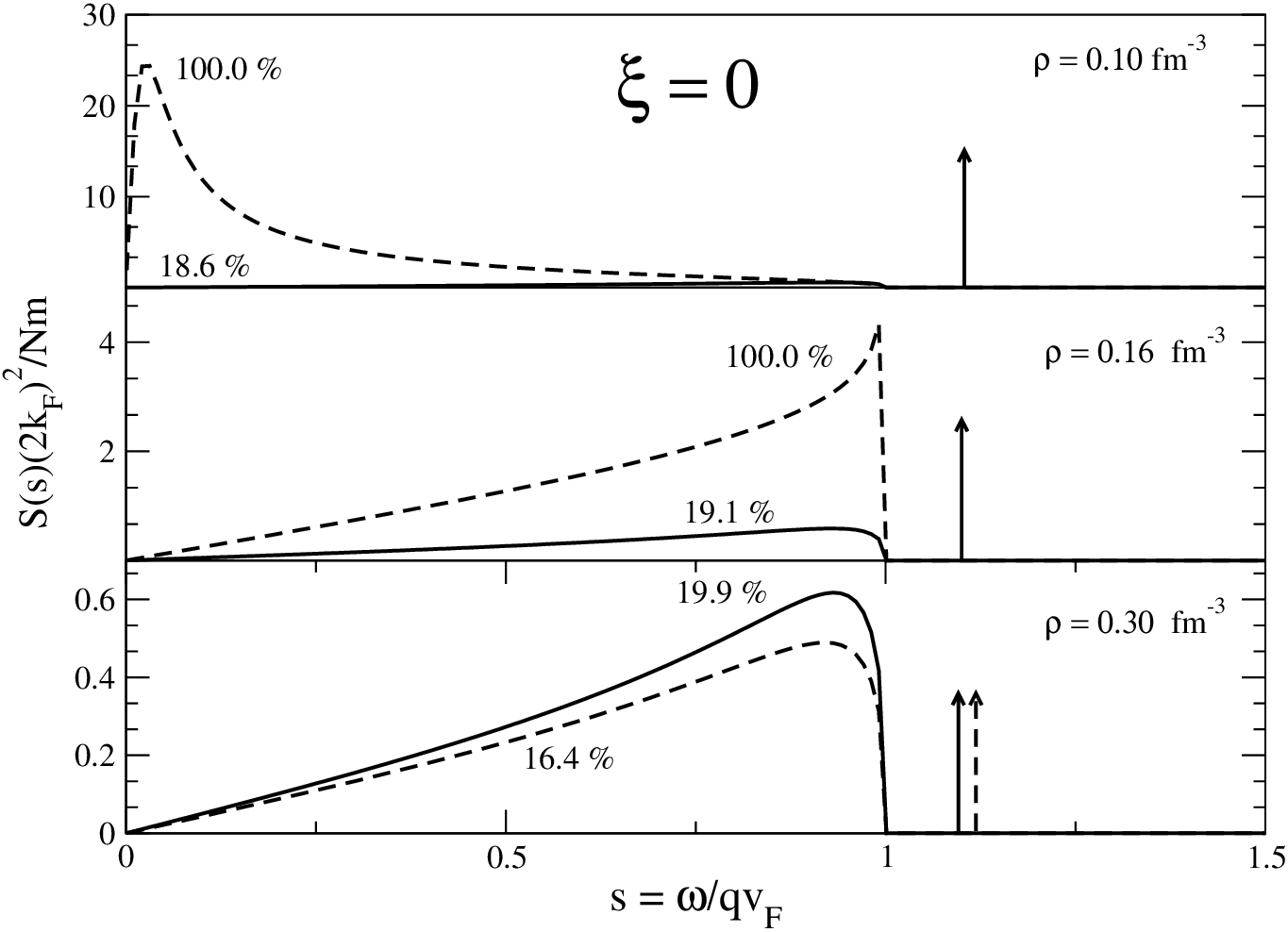}}

\caption[]{Quasi-particle-quasi-hole excitation strengths in units of $N m/2k_F^2$ as a function of $s=\omega/q v_F$ for three
different values of the density ($\rho=0.30 fm^{-3}$, $\rho=0.16 fm^{-3}$, $\rho=0.10 fm^{-3}$) at 
fixed isospin asymmetry $\xi=0$. The positions of the collective states, when present,  are indicated by an arrow.  
The dashed and full lines stand for the isoscalar and isovector excitation strengths, respectively. The fraction of energy-weighted
sum rule exhausted by the two strengths is also reported near the curves. The remaining fraction of energy-weighted sum rule is exhausted by
the collective states.}
\label{fig4}
\end{figure}

\begin{figure}{t}
 
\centerline{\includegraphics[scale=0.34]{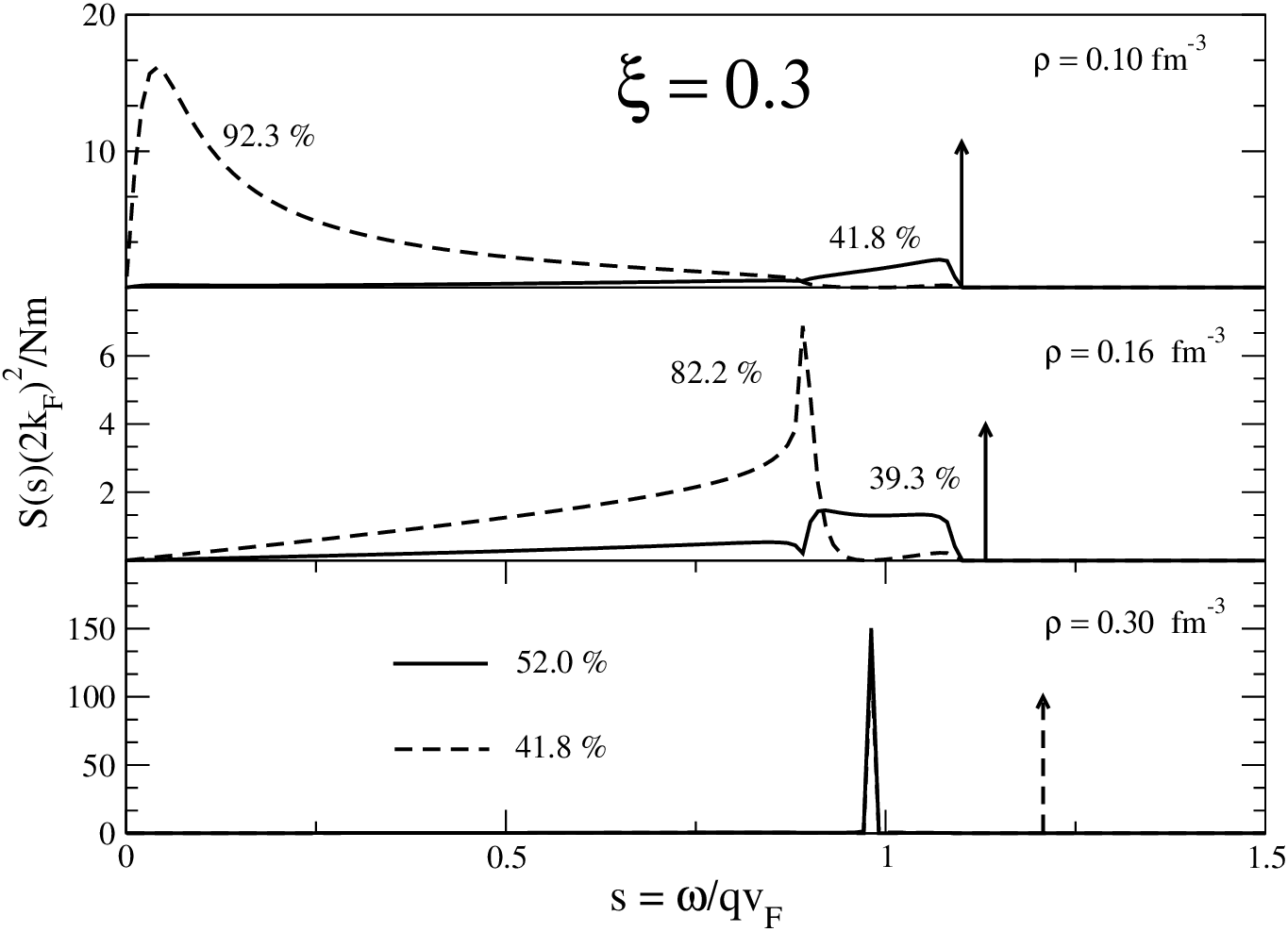}}
 
\caption[]{ Same as in Fig. 4 at
isospin asymmetry $\xi=0.3$. The collective state, indicated with an arrow, have the same energy both in the isoscalar and isovector channels.   
The dashed and full lines stand for the isoscalar and isovector excitation strengths, respectively. The fraction of energy-weighted  
sum rule exhausted by the two strengths is also reported near the curves. The arrow is dashed or full depending on whether the collective state exhausts more strength in the isoscalar or in the isovector channel.} 
\label{fig5}
\end{figure}

\begin{figure}{t}

\vspace{0.5cm}
\centerline{\includegraphics[scale=0.34]{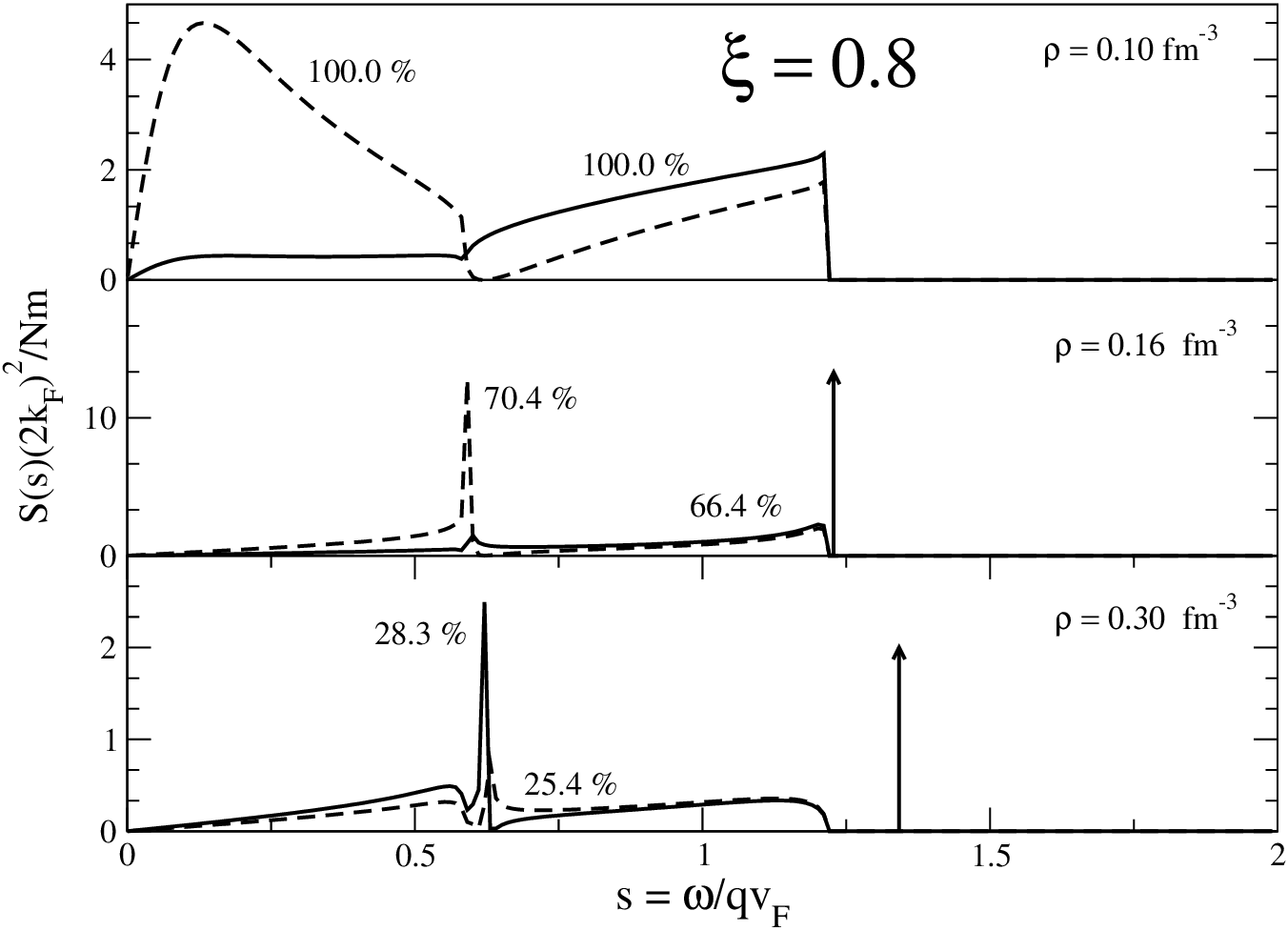}}

\caption[]{ Same as in Fig. 5 at
isospin asymmetry $\xi=0.8$.}
\label{fig6}
\end{figure}

\begin{figure}{f}

\vspace{0.3cm}
\centerline{\includegraphics[scale=0.34]{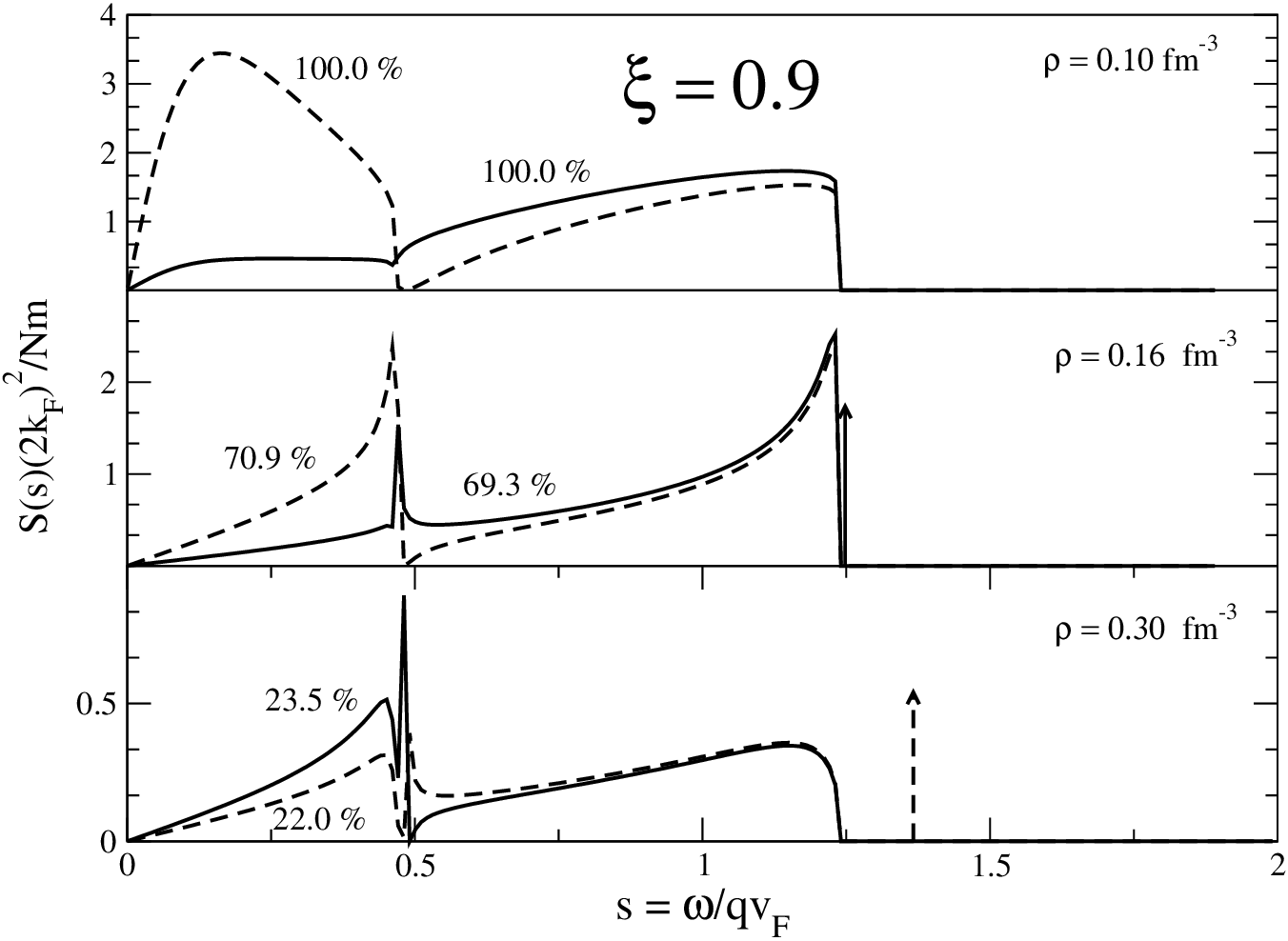}}

\caption[]{ Same as in Fig. 5 at
isospin asymmetry $\xi=0.9$.}
\label{fig7}
\end{figure}

\begin{figure}{t}

\centerline{\includegraphics[scale=0.34]{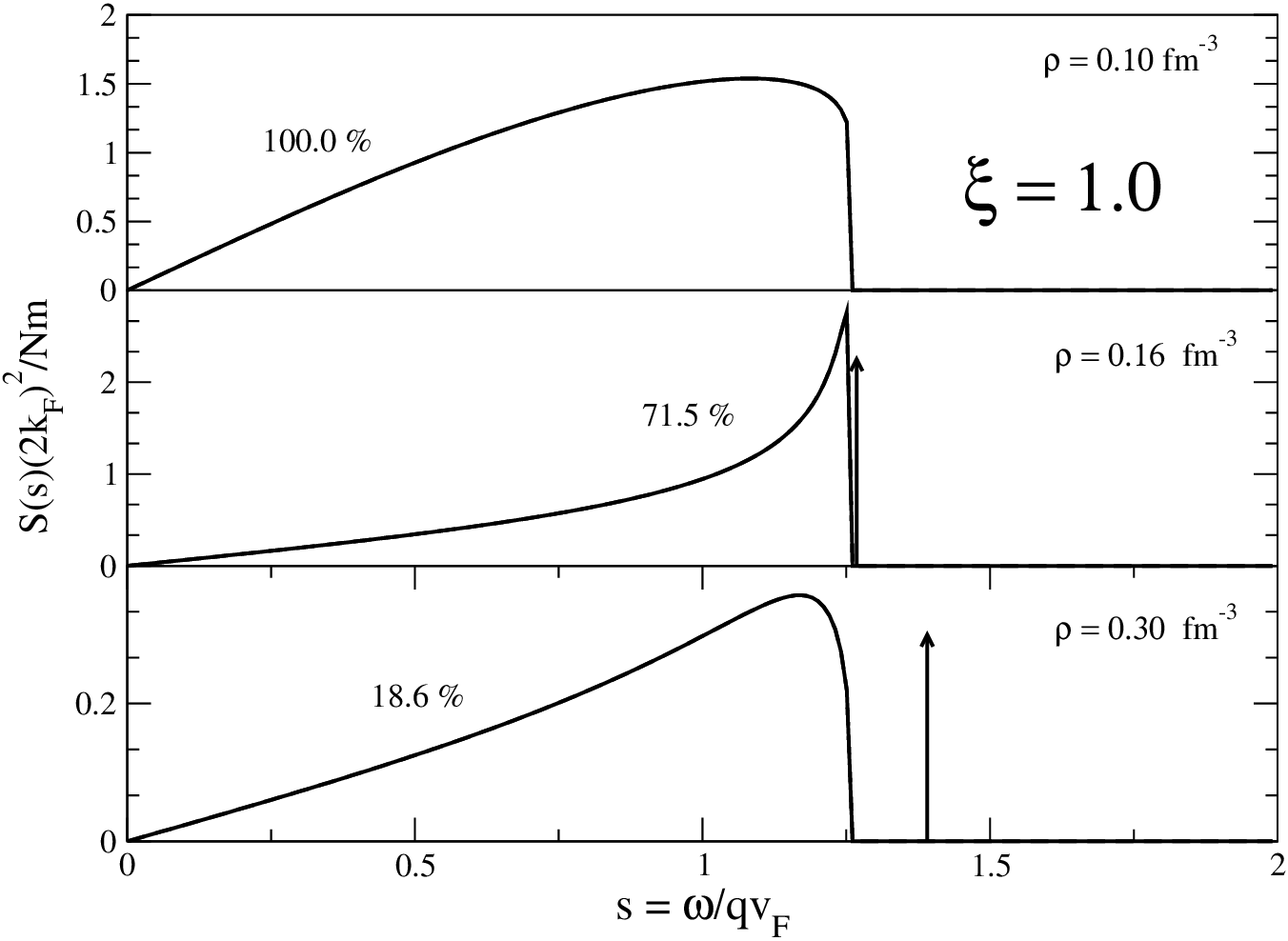}}

\caption[]{ Same as in Fig. 5 at
isospin asymmetry $\xi=1$. In this case, pure neutron matter,  there is only one type of excitation and the distinction between isoscalar and isovector channel does not apply.}
\label{fig8}
\end{figure}

\begin{table*}
\begin{ruledtabular}
\begin{tabular}{cccccccccccc}
$\xi$&$\bar{s}_s$&\% $m_1^s$&$\bar{s}_v$&\% $m_1^v$&$s^s_{1,-1}$&$s^v_{1,-1}$&$s^s_{3,1}$&$s^v_{3,1}$&$G_p$&$G_n$&$G_{np}$\\
\hline
0.0 &  -  &  -   & 1.1036 & 81.4 & 0.115  & 0.915 & 0.529 & 1.051  & 0.2757  &  0.2757 &  -1.2363 \\
0.3 & 1.1154  &  7.7   & 1.1154 & 58.2 & 0.142  & 0.925 & 0.472 & 1.088  & 0.3991  &  0.1422 &  -1.2031 \\
0.5 & 1.1466  &  5.5  & 1.1466 & 15.1 & 0.183  & 0.890 & 0.472 & 1.131  & 0.4611  &  0.0561 &  -1.1409 \\
0.8 &  -  &   -  &   -    &  -   & 0.269  & 0.891 & 0.564 & 1.105  & 0.4690  & -0.0599 &  -0.9578  \\
0.9 &  -  &   -  &   -    &  -   & 0.313  & 0.915 & 0.632 & 1.129  & 0.4093  & -0.0937 &  -0.8408  \\
1.0 &  -  &   -  &   -    &  -   & 0.681  & 0.681 & 0.942 & 0.942  & 0.0000  & -0.1247 &   0.0000  \\
\end{tabular}
\end{ruledtabular}
\caption{Numerical results at density $\rho=0.10$fm$^{-3}$ and various isospin polarizations for: a) the energy $\bar s$ of the collective mode in the isoscalar ($s$) or isovector ($v$) channels, and corresponding fraction of the $f$-sum rule exhausted by such modes; b) the centroids of the collective peaks excitations predicted by the energy-weighted sum rules (see Eq. (34)); c) Interaction parameters giving the mean-field potential.}
\end{table*}

\begin{table*}
\begin{ruledtabular}
\begin{tabular}{cccccccccccc}
$\xi$&$\bar{s}_s$&\% $m_1^s$&$\bar{s}_v$&\% $m_1^v$&$s^s_{1,-1}$&$s^v_{1,-1}$&$s^s_{3,1}$&$s^v_{3,1}$&$G_p$&$G_n$&$G_{np}$\\
\hline
0.0 &  -  &  -    & 1.1008 & 80.9 &  0.609  & 0.911 &  0.799 & 1.047 &  0.8012  &  0.8012 &  -0.6877 \\
0.3 & 1.1316   &  17.8 & 1.1316 & 60.7 &  0.614  & 0.909 &  0.793 & 1.076 &  0.8759  &  0.7077 &  -0.6610 \\
0.5 & 1.1685  &  26.8 & 1.1685 & 46.9 &  0.623  & 0.904 &  0.820 & 1.102 &  0.8953  &  0.6464 &  -0.6125 \\
0.8 & 1.2281  &  29.6 & 1.2281 & 33.6 &  0.650  & 0.901 &  0.916 & 1.150 &  0.7997  &  0.5669 &  -0.4826 \\
0.9 & 1.2481  &  29.1 & 1.2481 & 30.7 &  0.667  & 0.908 &  0.967 & 1.167 &  0.6751  &  0.5455 &  -0.4106  \\
1.0 & 1.2680 -  &  28.5 & 1.2680 & 28.5 &  0.899  & 0.899 &  1.110 & 1.110 &  0.0000  &  0.5271 &   0.0000  \\
\end{tabular}
\end{ruledtabular}
\caption{Same as table II for density $\rho = 0.16$fm$^{-3}$}
\end{table*}

\begin{table*}
\begin{ruledtabular}
\begin{tabular}{cccccccccccc}
$\xi$&$\bar{s}_s$&\% $m_1^s$&$\bar{s}_v$&\% $m_1^v$&$s^s_{1,-1}$&$s^v_{1,-1}$&$s^s_{3,1}$&$s^v_{3,1}$&$G_p$&$G_n$&$G_{np}$\\
\hline
0.0 & 1.1188 &  83.6 & 1.0958 & 80.1 &  0.937  & 0.903 &  1.070 & 1.040 &  1.5414  &  1.5414 &  0.0935 \\
0.3 & 1.2072 &  58.2 & 1.2072   & 48.0 &  0.943  & 0.902 &  1.099 & 1.059 &  1.5260  &  1.5267 &  0.1091 \\
0.5 & 1.2643 &  64.3 & 1.2643 & 57.8 &  0.953  & 0.900 &  1.148 & 1.094 &  1.4732  &  1.5146 &  0.1351 \\
0.8 & 1.3420 &  74.6 & 1.3420 & 71.7 &  0.986  & 0.902 &  1.263 & 1.188 &  1.2229  &  1.5064 &  0.1829 \\
0.9 & 1.3666 &  78.0 & 1.3666  & 76.5 &  1.011  & 0.909 &  1.310 & 1.232 &  1.0106  &  1.5084 &  0.1893 \\
1.0 & 1.3907 &  81.4 & 1.3907 & 81.4 &  1.153  & 1.153 &  1.324 & 1.324 &  0.0000  &  1.5134 &  0.0000  \\
\end{tabular}
\end{ruledtabular}
\caption{Same as table II for density $\rho = 0.30$fm$^{-3}$}
\end{table*}

In Figs.4-8 we then plot the  quasi-particle-quasi-hole excitation strengths $S^{s,v}(s=\omega/q v_F)$ in units of $N m/(2 k_F^2)$ for 
different 
values of the density $\rho$ and $s$ at fixed values of the isospin polarization $\xi$. This quantity has been obtained by numerically computing the immaginary part of expression 
(\ref{eq 21}). In the figures the fraction of energy weighted sum rule exhausted by 
the continuum of quasi-single particle states is explicitly indicated, and the position of the collective state, if present, is indicated by an arrow. 
In all the calculations presented in the following,
we have numerically checked that the particle-hole and collective contributions to the strength completly exhaust the $m_1$ sum rule.
Let us start the discussion of the two extreme cases corresponding to $\xi=0$ (symmetric nuclear matter $N=Z$) and $\xi=1$ (pure neutron matter
$Z=0$). In the first case ($N=Z$), isoscalar and isovector modes are decoupled in the sense that  the isoscalar and isovector density operators $F^s$ 
and $F^v$ give rise to distinct
isoscalar and isovector strengths $S^s(s=\omega/q v_F)$ and $S^v(s=\omega/q v_F)$ . This is different from what happens 
when $N\ne Z$ where $F^s$
and $F^v$  can indifferently excite modes both in the isoscalar and in the isovector channels, with strengths and weights depending on the values of $\xi$ and $\rho$.   
Obviously, for neutron matter ($Z=0$) it makes no sense to speak of isoscalar and isovector modes since in this case there is only one type of excitation.   

For symmetric nuclear matter ($\xi=0$) at density $\rho=0.30 $fm$^{-3}$,  in the isoscalar strength $S^s(s)$ a strong  
collective state  is present at ${\bar s}_s=1.1188$, exhausting about 83.6\%
of the energy weighted sum rule $m_1$. A continuum of single particle type
excitations is instead predicted at lower $s$ values. A similar situation occurs for $S^v(s)$. Here the collective state is at  at ${\bar s}_v=1.0958$, and 
exhausts about 80.1\%  of the energy weighted sum rule $m_1$. The collective states in both the cases occurr at energies which are 
closer to the mean excitation energies $\omega^{s,v}_{3,1}$ than to the $\omega^{s,v}_{1,-1}$ ones, showing that the collective modes are
of mainly of elastic type. This situation is analogous to what happens, for instance, in liquid He$^3$.
By decreasing the density, we still predict the existence of a strong 
collective state in the isovector channel, exhausting  about 80\%  of the energy weighted sum rule $m_1$. On the other hand, the collective mode disppears    
in the isoscalar channel, due to Landau damping.  

The situation is similar, but not exactly the same, in the case of pure neutron matter ($\xi=1$). At high density ($\rho=0.3$ fm$^{-3}$) the dynamic form factor shows a strong collective state of elastic nature at $\bar s=1.3907$, exhausting about 81.4\%  of the energy weighted sum rule, together with a continuum of single particle excitations at lower $s$ values.  
Decreasing the density to $\rho=0.16$ fm$^{-3}$, the energy of the 
collective state decreases, likewise the percentage 
of $m_1$ sum rule exhausted by this state. Eventually, at some density below  $\rho=0.16$ fm$^{-3}$, the collective completely decays in one particle-one hole excitations.  

When $N\ne Z$, isoscalar and isovector probes excite an admixture of isoscalar and isovector modes. In all cases we predict a single collective state, when  present, 
having different weights in $m_1$ for the isovector and isoscalar channels.

The typical situation expected in the neutron star core is that of a   
small quantity of protons present in the 
neutron matter, corresponding to $\xi$ close to 1.     
Rather than considering the proton fraction predicted for each value of the 
density by the AFDMC equation of state, we have considered 
the cases $\xi=0.8$ and $\xi=0.9$ for all densities, in order to have an idea of
the evolution of the collective modes with this parameter.
For such values of the asymmetry, a clean  collective state in both channels is present only at high density, where it exhausts about 70-80\% of the $m_1$ sum rule.
Already at saturation density $\rho=0.16$fm$^{-3}$,   
this state has almost completly  decayed into particle-hole excitations, since it exhausts only about the 30\% of the $m_1$ sum rule in both channels.
At $\rho=0.10 fm^{-3}$ it is completly damped.
This suggests that collective modes in the isoscalar or  isovectors channels can be present going inward from the interface between the inner crust and the outer core.

Another interesting analysis can be made by looking at the compressibility (in unit of the free one) as a function of the density for different values of the proton fraction. This quantity is given by $m^s_{-1}/ m^0_{-1}$ and $m^v_{-1}/ 
m^0_{-1}$, and is plotted in Figs. 9.  
It can be noticed that a strong divergence for a density near to $\rho=0.085 $fm$^{-3}$ appears at $\xi=0.9$, and that it moves to a larger density increasing the proton fraction. We interpret this divergence as a sort of mechanical instability of matter towards the formation of an inhomogeneous phase, as expected in the inner crust of the neutron star. 

It is also interesting to look at the regime that would correspond to asymmetries typical of very neutron rich nuclei ($\xi=0.3$). 
At (unphysical) high density, a collective state exhausting about the 50-60\% of $m_1$ at $\bar s=1.2072$ is present both in $S^s(s)$ and $S^v(s)$. 
At the same time the quasi-particle-quasi-hole strength is practically concentrated in one state at a value of $s$ sligthly 
smaller than 1 and exhausting about the 50-40\% of $m_1$.
By lowering the density, we observe that 
at $\rho=0.16$fm$^{-3}$ the collective state appears at a lower energy, and remains collective only in the isovector channel, where it still exhausta more than 60\% of theenergy weighted sum rule.  In the isoscalar channel most of the strength is taken by the one-particle-one excitations, though concentrated in a narrow region of $s$, and the "collective" solution exhausts only about 17.8\% of the  $m_1$ sum rule. At $\rho=0.10 $fm$^{-3}$ the collective mode can be excited 
only by an isovector probe, and in the isoscalar channel it is practically completly damped. The quasi-particle-quasi-hole strength is distributed in a wide 
range of $s$ in the isoscalar channel, whereas in the isovector one continues to see a substantial concentration at  $s\sim1$.

\begin{figure}{t}

\centerline{\includegraphics[scale=0.34]{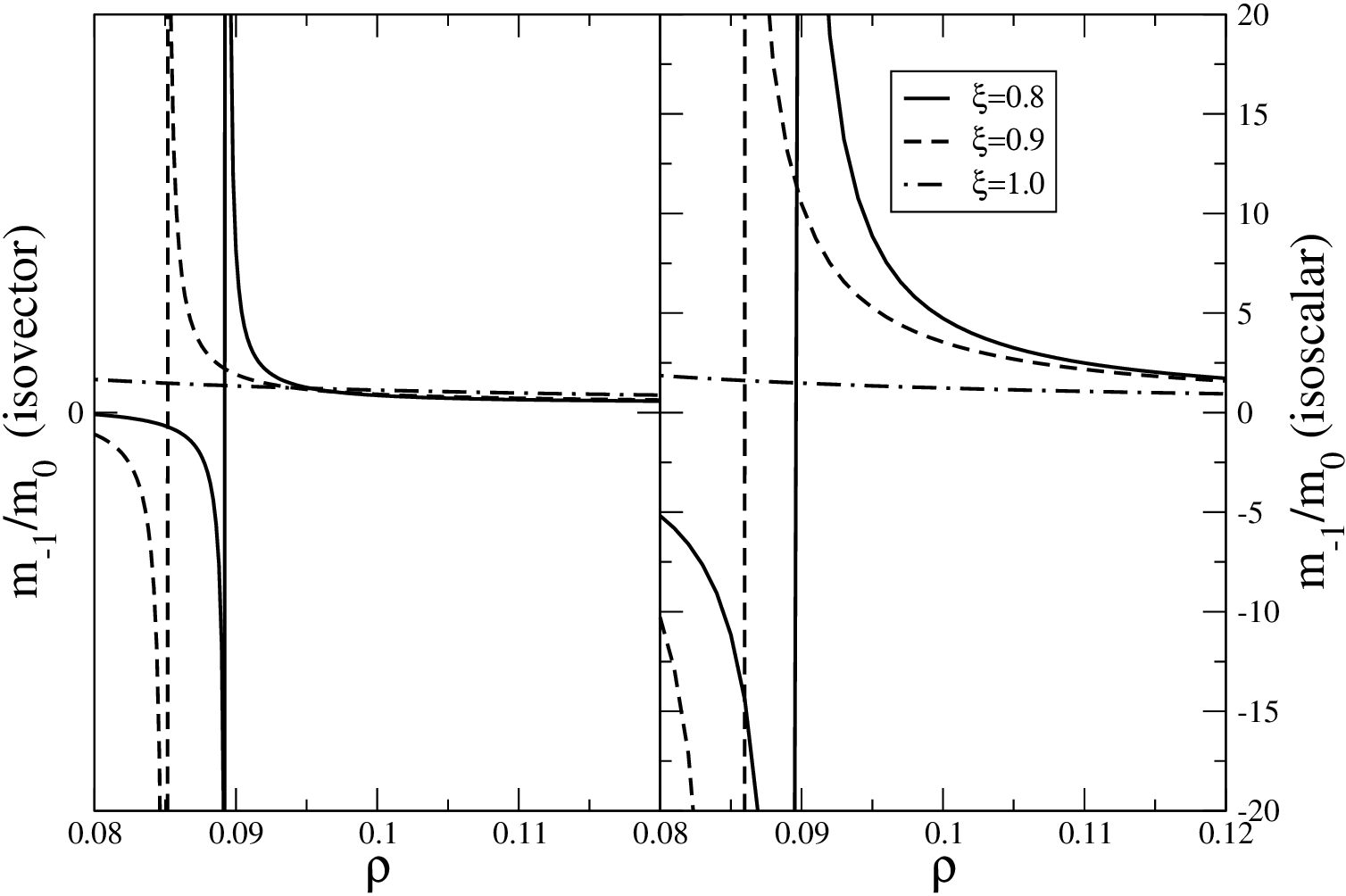}}

\caption[]{Isoscalar and isovector compressibility ratios   $m^s_{-1}/ m^0_{-1}$ and $m^v_{-1}/ m^0_{-1}$
as a function of the density $\rho$ for values of $\xi$ close to 1.}
\label{fig9}
\end{figure}

This situation might be interpreted in analogy to what happens in nuclei with a large excess of neutrons, where a low energy peak beyond the usual giant
resonance is observed in the isovector channel. 

\begin{figure}{b}

\vspace{0.5cm}
\centerline{\includegraphics[scale=0.34]{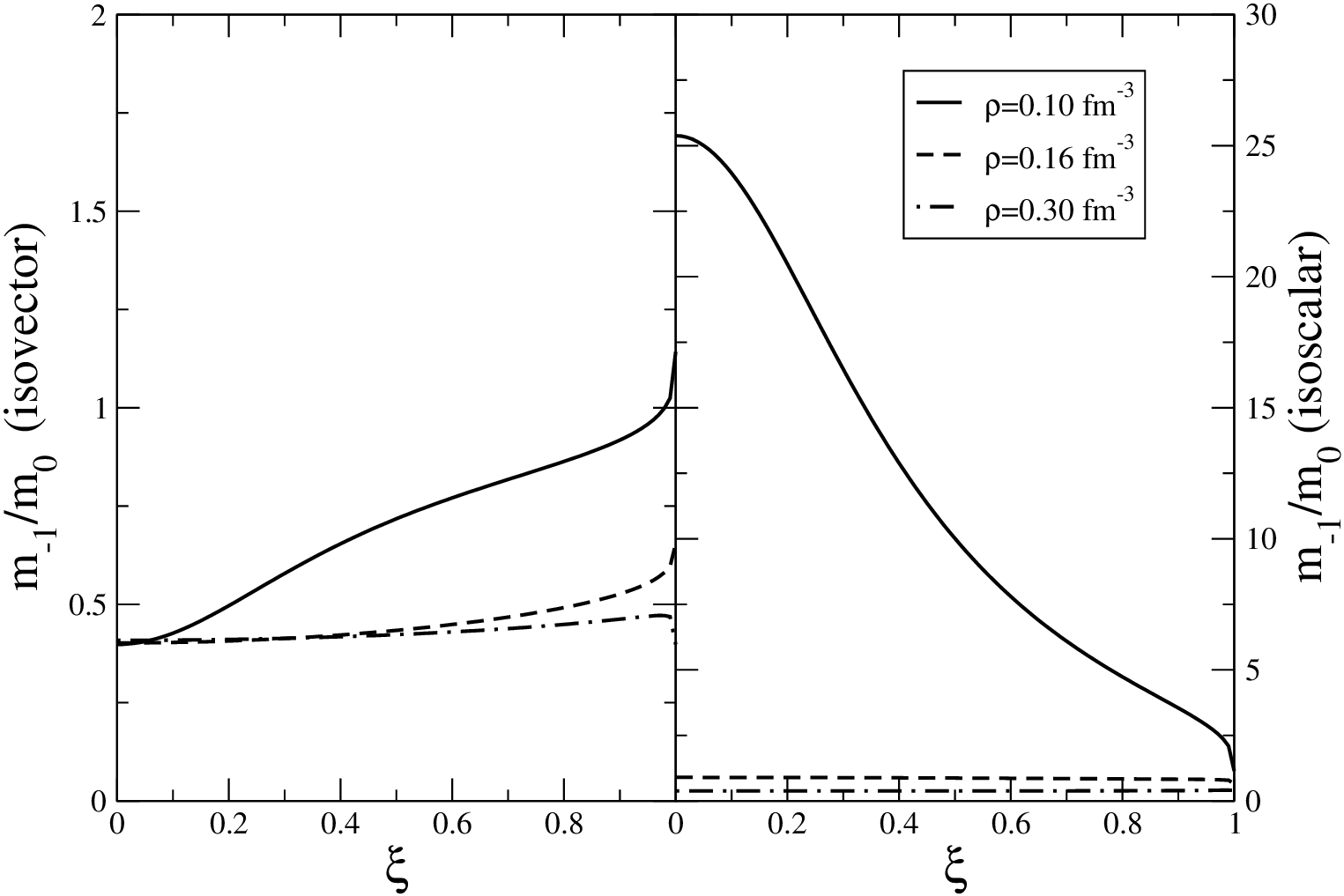}}

\caption[]{Isoscalar and isovector compressibility ratios $m^s_{-1}/ m^0_{-1}$ and $m^v_{-1}/ m^0_{-1}$
as a function  of the isospin asymmetry $\xi$  for three different values of the density.
}
\label{fig10}
\end{figure}

Finally, in Figs. 10 we plot the compressibility ratio for three different values of the density. At $\xi=0$ and $\xi=1$, $m^s_{-1}/ m^0_{-1}$ gives the compressibility of symmetric nuclear matter and  pure neutron matter, respectively. These values are not far from the  results computed by other authors starting from a microscopic AV8' Hamiltonian\cite{Fan01}. From the figures one also sees that 
differently from what happens at low density, at normal and high densities,
$m^s_{-1}/ m^0_{-1}$ is practically independent on $\xi$. On the contrary,  $m^v_{-1}/m^0_{-1}$ results to be always an increasing function 
of the isospin asymmetry.

\section{Conclusions}

We have studied the dynamic form factor of asymmetric nuclear matter by using a time-dependent local isospin density approximation 
approach based on a local density energy functional derived by an Auxiliary Field Diffusion Monte Carlo calculation. 
The more relevant results we have found are the following:
i) The presence of a strong collective state at high density at an energy which increases with the value of the nuclear asymmetry $\xi=(N-Z)/A$; 
decreasing the density down to saturation ($\rho=0.16$ fm$^{-3}$), this state tends to decay in quasi-particle-quasi hole states. The decay is faster  in the isoscalar channel, where it remains very collective
only at values of $\xi$ close to 0. 
Further decreasing the density, the collective states survives only in the isovector channel at small values of $\xi$, which are typical of the neutron star interior.
ii) When a small fraction of protons is added to the neutron matter and $\xi$ is equal to 0.8, 0.9, at values of the density near or slightly smaller of 
0.09 $fm^{-3}$ the system becomes unstable. This unstability is seen in the isoscalar and
isovector compressibilities, which at such densities diverge.
iii) When $\xi$ is around 0.3 (small asymmetry), in the isovector channel at all the densities two states cohexist, one of collective and
the other of quasi-particle-quasi-hole nature, practically sharing the fraction of exhausted energy-weighted sum rule. This
is similar to what is observed in the photodisintegration of large neutron excess nuclei.     

\section*{ACKNOWLEDGMENTS}
We thank W. Leidemann and G. Orlandini for useful discussions about the subject of the paper. FP is a member of LISC, Interdisciplinary Laboratory for Computational Science, a joint venture of the University of Trento and of the Bruno Kessler Foundation.


\end{document}